# A Survey of Neighbourhood Construction Models for Categorizing Data Points

Shahin Pourbahrami[1,*], Leyli Mohammad Khanli[1]

*Computer Engineering Department, Faculty of Electrical and Computer Engineering, University of Tabriz, Tabriz, Iran1*

## Abstract

Finding neighbourhood structures is very useful in extracting valuable relationships among data samples. This paper presents a survey of recent neighborhood construction algorithms for pattern clustering and classifying data points. Extracting neighborhoods and connections among the points is extremely useful for clustering and classifying the data. Many applications such as detecting social network communities, bundling related edges, and solving location and routing problems all indicate the usefulness of this problem. Finding data point neighbourhood in data mining and pattern recognition should generally improve knowledge extraction from databases. Several algorithms of data point neighbourhood construction have been proposed to analyse the data in this sense. They will be described and discussed from different aspects in this paper. Finally, the future challenges concerning the title of the present paper will be outlined.

## 1. Introduction

A neighbourhood includes a group of data points which can locally co-occur and are defined according to their local interconnectivity in a database (İnkaya et al. 2015, O'Callaghan 1975). One of the important features in data mining and pattern recognition is the analysis of the data and their categorization into similar groups through precise examination of neighbourhood features of the data. In other words, the neigbourhood for each set of data is defined in terms of similarity and connection among the data so that the obtained accuracy can respond to the present demand for data mining in different fields (Zhou et al. 2005, Ahuja et al. 2002). Neighbourhood can provide a base for making accurate models according to the similarities among samples by grouping the data into similar categories. In other words, the purpose is to discover the hidden and implicit information as well as the interconnections in the current data and to predict the ambiguous or unobserved cases.

Nowadays, with increasing human information, using algorithms such as data analysis for extracting knowledge is inevitable. Due to high amount of data in many applications and the growing significance of the new data sets, it is not economical to save data. Therefore, the data to be processed always change dynamically. Hence, processing and classifying the data in different fields such as social networks and clustering of data points for detecting community structures (İnkaya 2015, Singh et al. 2016, Zhou et al. 2017, Wang et al. 2016, Bhattacharya et al. 1981, Wang et al. 2017) data visualization (Arleo et al.. 2017, Guo et al. 2017) and clustering (Von Luxburg 2007, Maier et al. 2013, Chen et al. 2011, Ester et al. 1996, Breunig et al. 2000, Jain 2010, Sander et al. 1998, Ertöz et al. 2003, Iyigun 2008, Pourbahrami 2017, Pourbahrami et al. 2018), as well as classifying data in many scientific and economic fields is inevitably necessary.

In this paper, we examine the studies and algorithms in the literature of neighbourhood construction, taking into account the similarity among the data sets so that useful knowledge would be extracted. The studies on the clustering of data flows (based on neighbourhood construction) will also be examined. Neighbourhood construction algorithms may be grouped into four major categories based on their principles and their underlying structures:

[*] Corresponding Author.

*Email Addresses:* sh.pourbahrami@tabrizu.ac.ir (Shahin Pourbahrami), l-khanli@tabrizu.ac.ir (Leyli Mohammad Khanli).

(1) Algorithms based on distance or parametric approaches: Algorithms with *k*-Nearest Neighbor and *ε*-neighbourhood use *k* or *ε* to construct neighbourhood structures.

(2) Algorithms based on the principle of geometric structures where the geometric shape is formed at neighbourhood location.

(3) Algorithms based on artificial neural network which use competitive learning algorithms.

(4) Algorithms based on the graph and density structure which use mutual neighbourhood information or the distance between the nodes.

The rest of the paper is organized as follows. Section 2 will describe the distance-based approaches for neighbourhood construction. In Section 3, we will present the geometric and graph-based approaches, and in Section 4 the approaches based on neural networks will be presented. Section 5 will present a summary of the graph and density-based approaches. Section 6 will describe the challenges and directions for neighbourhood construction approaches. Finally, Section 7 concludes the paper and provides the future challenges in this field.

## 2. Neighbourhood Construction: Distance-based and Density-based Approaches

One of the most well-known algorithms that used to define and cluster neighbourhood data (i.e., graph-based data and mutual information based data) using Euclidean distance is *k*-Nearest Neighbour (*k*-NN). According to Qin et al. (2018), k-NN starts with choosing the data sets that are similar to each other. Later, it determines the clusters. The mentioned algorithm is very simple and is highly useful in determining neighbourhood. Moreover, it is the only criterion used for determining neighbourhood by the use of *k*-NN algorithms, the distance as well as place of the points. Moreover, the statistical rules and the geometric structures are not taken into account. In Fig. 1, *k*-Nearest Neighbor is shown; in this figure, 3-nearest neighbours for *i* are points *j*, *m*, and *n*. It means that the set of *i* for *k*-nearest neighbours is the set of *k* ($k$>0) nearest neighbours for *i*, and is indicated by *k*-NN (*i*) (İnkaya et al. 2015, Von Luxburg 2007, Duda et al. 2012, Lu and Fu 1978, Jarvis and Patrick 1973, Tao et al. 2006).

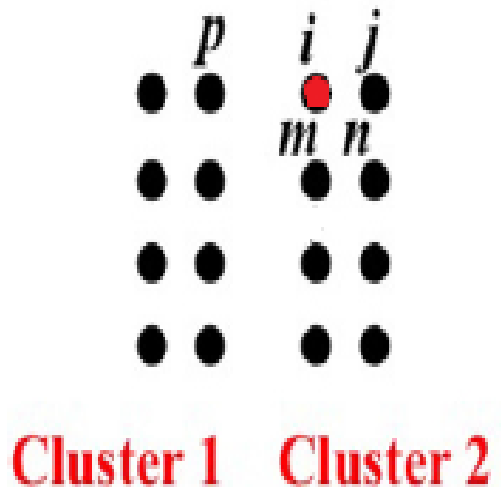

**Fig. 1**. k-NN neighbourhood (İnkaya et al. 2015).

Density-based clustering is a clustering method in which the clusters are defined as an area with higher density than that of the other samples. In other words, this method of clustering is based on the premise that the clusters include an area of high density data space separated by less dense areas. A cluster (as a connected density element) will grow in a direction causing density or creating dense areas. Therefore, these algorithms are able to identify the clusters with different shapes; they will also provide a natural support against the outlier data. The most common algorithms in this group are: DBSCAN (Ester et al. 1996), OPTICS (Ankerst et al. 1999) , DBCLASD (Xu et al. 1998), GDBSCAN (Sander et al. 1998), DENCLU (Hinneburg and Keim 1998), and DPC (Rodriguez and Laio 2014), DBSCAN: Density-based clustering has been proposed by DBSCAN algorithm. In this method, each piece of data belongs to an algorithm which is available to other data densities in the same cluster, but it is not available for the data densities in other clusters (Ester et al. 1996). This type of clustering has some advantages. For example, it is not sensitive to the shape of the data, can identify the clusters with various non-spherical shapes in the data, the number of clusters is automatically determined simultaneously with clustering, and it is efficient in identifying noise. One of the main disadvantages of this algorithm may lie in its unpredictability for borderline behavior samples. Neighbourhood is determined by the simple epsilon (ε) based on two factors, namely small neighbourhood radius and epsilon distance (Von Luxburg 2007, Pedrycz 2010, Saha and Bandyopadhyay 2009). Data point labels are determined in this phase of neighbourhood construction. The selection of inappropriate parameters will cause a decline in the efficiency epsilon in this approach. As a result, it seems that regarding neighbourhood structure, this algorithm would not be as strong and accurate as other algorithms. If the value reported for the epsilon in this algorithm is small, there is a strong tendency for the point under focus in the respective radius to have no neighbours; as a result, the point may be taken as the outlier data, although this is not really the case. In case of having wide neighbourhood radius, it is likely that the different points from different groups may integrate into each other and form a single group. In this method, clusters have been grouped into three distinct categories including: 1. Core samples, 2. Border samples, 3. Noise samples. A type of ε-neighbourhood has been displayed in Fig. 2. In this Figure, when the values for epsilons are small, *r* is likely to be located in outlier point. Moreover, in this type of neighbourhood the clusters with different density rates are not readily available, and they fail to work with more than one outlying point (Tsai and Huang 2009).

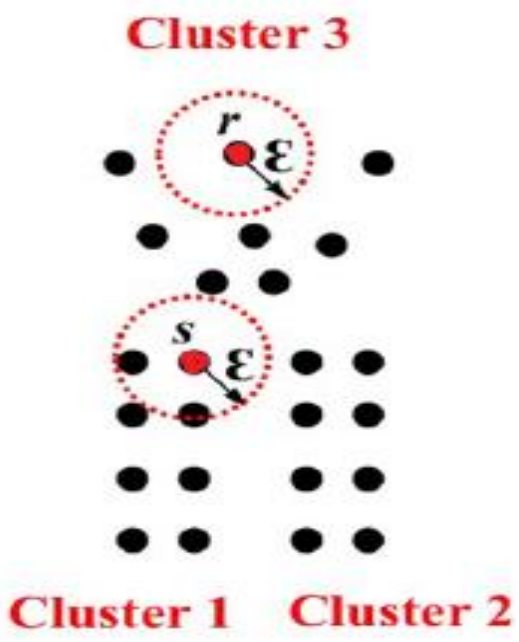

**Fig. 2**. ε-neighbourhood (İnkaya et al.. 2015).

DENCLU: It is one of the other density-based algorithms which based on density function (Hinneburg and Keim 1998). It examines the data sets in the network structure and finds dense cells according to their means. It also finds density absorbents using hill climbing algorithm; these absorbents should have the maximum local amounts all over density function. Finally, it integrates the absorbents to produce the final clusters. This algorithm has better performance in big data sets and identifies the clusters with all possible shapes.The other advantages include: their ability to identify clusters (in different shapes) and noise, as well as the number of clusters automatically simultaneous with the act of clustering.

Disadvantages: The efficiency of these methods largely depends on adjusting their parameters indicating their over sensitivity to parameters (only some of the parameters of this method suffer from this problem); such algorithms may fail to show good performance in borderline and overlapping cases. The mentioned problems, however, have been gradually resolved in this method. Meanwhile, some problems still persist, and it is possible for locating and identifying better algorithms.

OPTICS is an improvement of ε-neighbourhood, and it overcomes the shortcoming of ε-neighbourhood such as being sensitive to two parameters, the radius of the neighbourhood, and the minimum number of points in a neighbourhood (Ankerst et al. 1999). In the process of Mean-shift, the mean of offset of current data point is calculated at first, the next data point is figured out based on the current data point and the offset. Finally, the iteration will continue until some criteria are met (Ankerst et al. 1999, Januzaj et al. 2004, Kriegel and Pfeifle 2005, Chen and Tu 2007, Duan et al. 2007). The main idea behind this algorithm is similar to that of DBSCAN, but it deals with the main weaknesses of DBSCAN. In fact, it accounts for the weakness of this algorithm in identifying meaningful clusters in the data sets with different densities. It works on the basis of cosinus distances among the samples, and to form a new cluster the number of the samples in a data set should exceed a given number, which is by itself one of the algorithm parameters. In this method, the last cluster separates the residual samples from the other clusters. For this reason, in most of the cases, the size of the last sample exceeds that of the others.

Any clustering or classification of data points involves integrating or identifying objects that are near or similar to each other. Distances metrics or similarities metrics are mathematical representations of what we mean by close or similar. The choice of distance is extremely important. In some cases, a Euclidean metric will be sensible while in others a Manhattan metric will be a better choice. Generally, some experience or subject matter knowledge is very helpful in selecting an appropriate distance for a given project. The commonly used distance metrics for quantitative data set feature are summarized in Table 1.

**Table 1. Distance Metrics for Categorizing Data Points**

| Distance Metric | Formula | Type of Attributes | Time complexity | Advantage | Disadvantage |
|---|---|---|---|---|---|
| Euclidean Distance | $\left(\sum_{l=1}^{d}\left\|\frac{x_{il}-x_{jl}}{s_l}\right\|^2\right)^{1/2}$ | Numerical attributes | O (n) | • Very common and easy to compute and works well with data sets with compact | • Sensitive to outliers |
| Value Difference Metric (VDM) | $d_{vdm,a}(x,y) = \sum_{c=1}^{C}\left\|P_{x,c}-P_{y,c}\right\|^q$ | Symbolic/nominal attributes | O (n) | • Uses the mutual information between the attribute<br>• Handle continuous attributes | • Inappropriate to use directly on continuous attributes |
| Minkowski Distance | $\left(\sum_{l=1}^{d}\left\|x_{il}-x_{jl}\right\|^n\right)^{1/n}$ | Ordinal and Quantitative attributes | O (n) | • Generalization of both the Euclidean and the Manhattan<br>• Implemented in spatial analytical modeling<br>• More reliable results | • The use of Exclusive Assignment<br>• The learning algorithm is not invariant to non-linear transformations |
| Cosine Distance | $1-\cos\alpha = \frac{x_i^T x_j}{\|x_i\|\|x_j\|}$ | Text attributes | O (3n) | • Independent of vector length and invariant to rotation | • Not invariant to linear transformation |
| Pearson Correlation Distance | $1-\frac{Cov(x_i,x_j)}{\sqrt{D(x_i)}\sqrt{D(x_j)}}$ | Random vectors | O (2n) | • Be used to isolate potential outliers | • Depending on the numbers involved |
| Mahalanobis Distance | $\sqrt{(x_i-x_j)^T S^{-1}(x_i-x_j)}$ | Random vectors | O (3n) | • Non degenerately<br>• Used to detect outliers | • Expensive in terms of computation |

Density peak Clustering (DPC) is a new density-based and distance-based algorithm which was introduced by Rodriguez and Live in a valid science journal in America (Rodriguez and Laio 2014). It is based on the approach that the cluster centers are obtained using high densities of their neighbors (local density) with a rather high distance from high density samples (Delta). For each sample, these quantities are calculated according to the matrix of sample distances (binary distance of the samples from each other). Unlike center-based algorithms, this algorithm does not need the parameter of the number of clusters and repetitive process. In fact, the best cluster samples are obtained using the two mentioned quantities. Its only parameter is cut-off distance which is used for determining neighbourhood percentage in calculating sample densities. After finding cluster centers, the remaining samples are allocated to cluster centers using the strategy of the nearest neighbor having the highest density. In other words, each sample takes the neighbourhood label with higher density than that of the sample. This method can identify the samples with non-convex shapes and can identify noise samples to some extent. This algorithm as well as its advantages, and disadvantages will be discussed in the following section. Calculating the local density ($\rho_i$) and Delta ($\delta_i$) which both depend on the density between the samples is expressed in the following way. The local density of sample is calculated using the following equation:

$$\rho_i = \sum_{j \neq i} X\,(d(x_i, x_j) - d_c) \tag{1}$$

Where $d(x_i.x_j)$ is the Euclidean distance between sample $x_i$ and $x_j$ , and $d_c$ is the cutoff distance. In fact, the process of choosing $d_c$ parameter pertains to choosing the mean number of neighbours in the data set. Moreover, $\chi\,(x) = 1$ if $x < 0$ and otherwise $\chi\,(x) = 0$.

Delta, which is the least distance between sample i and any other sample *j* with higher density, is expressed in the following way:

$$\delta_i = \min_{j:\rho_j > \rho_i} (d(x_i, x_j)) \tag{2}$$

In the points with the highest rate of density, the following equation is defined:

$$\delta_i = \max_j (d(x_i, x_j)) \tag{3}$$

After calculating DPCs of local density ($\rho_i$) and Delta ($\delta_i$), it is possible to locate cluster centers using them. This algorithm obtains cluster centers using a decision graph; it is a chart in which the horizontal axis is local density and its vertical axis is Delta. In this algorithm, the sample with higher density and relatively high delta is selected as the cluster center. In fact, none of the two mentioned quantities can indicate a suitable center by themselves. Therefore, in decision graphs, the samples with high density and delta are selected as cluster centers, and the samples in the right top corner of the decision graph are considered as cluster centers. After finding the clusters, every sample takes the label of the neighbour having the highest density. This label allocation strategy first starts with samples neighbouring the cluster center, and takes the label of their centers as the neighbour with the highest density rate. In this way, labels are allocated to all samples.

Advantages: Identification of samples with given shapes: The clusters with given shapes and labeling the samples are identified using the criterion of density, and it is one of the principle components of density-based methods to identify the clusters with given shapes. Lack of repetition phase: label allocation to cluster samples in cluster centers takes place in a single phase. Lack of the primary center selection parameter: By using the concept of local mass data and delta extracted from the samples, the samples with relatively common features are identified. In fact, the centers are extracted from the basic features of the samples. Not having residual parameters and having only a single parameter for calculating local density: The proposed algorithm has superior performance in comparison with other algorithms such as DBSCAN which needs two parameters for clustering. Moreover, arranging a parameter in large scale data sets would not be hard.

Disadvantages: The strategy of allocating one point to the nearest neighbour with higher density can cause wrong distribution in clustering. In other words, if a given sample takes the wrong label, this wrong process would be dispersed to the other samples and would affect clustering performance. Cut parameter on the data sets is helpful in calculating local density in a general way; therefore, the density of each sample as a local criterion may not serve as a suitable criterion. It may also bear a negative effect in large data sets and may cause cluster loss. The selection of centers on decision graph will be associated with user intervention. Moreover, on some data sets, the number of appropriate clusters obtained from the decision graph may not be equal and comparable with the number of ideal clusters.

One study (Du et al. 2016) examines the challenge of general structure of the data to calculate density; it also offers a strategy for higher performance in the data at a relatively wide scale. Moreover, it has analyzed the principal components for better performance of the algorithm for the purpose of reducing data dimensions. However, it has offered no idea for allocating labels to non-central samples. Furthermore, analyzing the principal components has failed to have a considerable performance in reducing dimensions of the problem. Based on the procedure of label allocation, this method would not show acceptable performance in overlapping data sets. Another research uses the idea of local density allocation for each sample (Lotfi et al. 2016) to determine cluster centers. Then, it appoints a neighbourhood for each cluster sample, and all of the samples inside each cluster sample take the label of that specific cluster center. For the remaining samples, a simple but quick strategy based on sample densities and voting among the samples is adopted. In fact, the labeled samples are distributed, and the unlabeled samples are also labeled. Despite the acceptable performance of this method, there would still be weaknesses in label distribution which affect the efficiency of this method. It has some disadvantages including its poor performance in cluster intervention and lack of any strategy in dealing with the outlier samples. After obtaining cluster samples, another study (Xie et al. 2016) has adopted two strategies for unlabeled samples. It has divided the data sets into two types including core data and outlier data. The first strategy which is applied to data cores adopts the same cluster sample for each neighbor of that sample, and if the neighbors of that cluster sample are less than a threshold level, they take the label of the same cluster sample. Data cores are labeled similarly and are placed in their respective clusters. The second strategy is adopted for the outliers. In fact, for the residual unlabeled samples, membership degree for each cluster center is calculated. This method has acceptable performance, but it suffers from some disadvantages such as failure to adopt appropriate strategy and efficiency to obtain outlier data, as well as its increased sensitivity towards neighbourhood decision parameter. The other weakness of this method lies in the fact that it does not differentiate between noise and border data, which may affect clustering performance, especially in noise and real life data. In the process of reaching appropriate efficiency, this method is highly costly in terms of time; this is especially the case with large scale data. It should also be explicated that this method does not have appropriate performance for the clusters having very complicated shapes. In fact, due to inappropriate single phase distribution of the data, it may have problems in overlapping clusters. Some other methods will be discussed in this regard some of which have some improvement but suffer from some disadvantages.

Another study (Yaohui et al. 2017) selects cluster centers comparatively. In fact, it locates a large number of centers in the data sets which outnumber the real cluster centers. Later, they integrate with each other according to the nearness of the clusters. The only advantage of this method is finding appropriate and suitable samples for cluster centers. However, its main disadvantage lies in lack of a suitable strategy for integrating the samples and obtaining real clusters. In some data sets, some clusters are integrated wrongly, which dramatically reduces clustering performance. There are still some other disadvantages with this algorithm including overall density calculation, the way of allocating sample labels and clusters having complicated shapes as well as noise samples. In another study (Wang and Song 2016), rather than using decision graph observations to identify cluster centers, a statistical test was used. The authors (Mehmood et al. 2016) have proposed a method using a temperature equation called CFSFDP-HD to obtain a better estimate of local density and reduce its sensitivity to the rate of cut parameter. In this method (Qiu and Li 2015), a hierarchical density-based clustering called as D-NND has been proposed which learns density structure and cluster structure simultaneously. It also prevents estimation of too soft density. It also shows some reliability level against parameter sensitivity. In another study (Zhang et al. 2017), maximum density clustering for clustering a large number of dynamic data in the internet has shown considerable improvement. In fact, its purpose is offering an incremental maximum density clustering which can improve clustering results effectively and efficiently with the arrival of the new data. Incremental clustering method reduces processing time which makes the algorithm more efficient for industrial use.

Still in another study (Wu et al. 2017), a clustering method for maximum grid-based density was offered. In this method, instead of using Euclidean distance among all samples, a limited number of grid samples are defined. Moreover, a new method has been used to calculate sample density. Due to the use of grid methods, this method has higher executive speed. Method (Shi et al. 2018) is based on density, and *K* calculates the nearest neighbourhood for all samples. Later statistical methods have been used to measure density or ensure distributional scattering of all samples. In another study, (Xu et al. 2015), a clustering method of maximum manifold density called MDPC has been proposed. In fact, geodesic space is map into manifold space. In other words, by using manifold-based method the data sets with high dimensions is mapped into lower dimensions. In another study (Liang and Chen 2016), divide and conquer strategy has been used to improve clustering method. This method automatically obtains the number of clusters without any need for user knowledge. Some other methods have used the idea of *K* nearest neighbourhood (e.g., research Mehmood et al. 2016) has proposed a comparative maximum density clustering based on nearest neighbourhood *K* with a comprehensive strategy.

Research (Zhang et al. 2016) has examined distributed efficiency algorithm for clustering maximum density so that it can reduce calculation costs. In fact, map reduce solution has been studied, which offers a method called as LSH-DDP; it is an approximate algorithm which extracts a local sensitive hashing of data areas, conducting local calculations, and accumulating local results. Another study (Xu et al. 2016) has offered a maximum clustering density method based on a hierarchical structure called Den PEHC. By introducing a network structure, clustering ability in data sets with high and big dimensions increases, therefore, some authors have offered a method called DPC-MD which generalizes a method called DPC-MID; it calculates maximum density clustering by the use of maximal density clustering using a mixed area. This type of clustering generalizes maximum density by the use of a general criterion for the integrated data.

## 3. Construction: Geometry-based and Graph-based Approaches

Graph and neighbourhood topology have also been utilized to enhance classification by means of some algorithms. Some graphs including Gabriel's graph neighbours are, in fact, geometric algorithms used to examine all types of relationships persisting among the points (İnkaya et al. 2015, Hashem et al. 2015, Guedes et al. 2016, Zahn 1970, Koontz et al. 1976, Urquhart 1982, Gabriel and Sokal 1969). Fig. 3 indicates that in Gabriel's graph the distance between point *a* and *b* is used as the circle diameter. Point *a* and *b* are considered to have a direct relationship with neighbourhood when no other point is left within Gabriel's neighbourhood circle. If this condition is not fulfilled, the number of points inside Gabriel's graph is selected as the density measure between the two points.

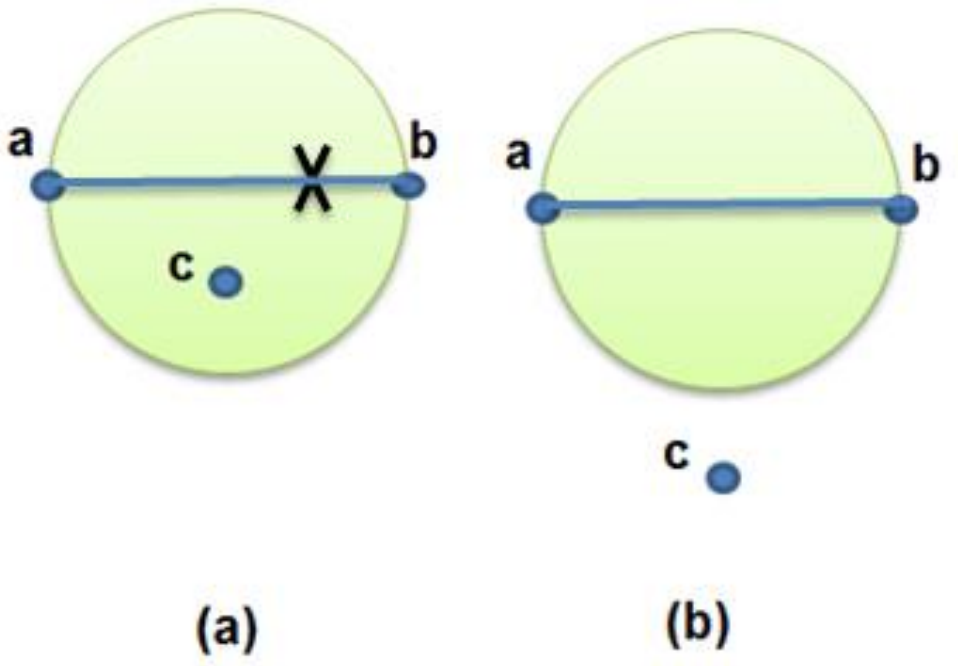

**Fig. 3**. Example neighbourhoods constructed with Gabriel graph . (a) indirect neighbourhood, and (b) direct neighbourhood.

The geometric structure and density between the points are used as the criteria for establishing neighbourhood using Gabriel's graph in NC algorithm (İnkaya et al. 2015). In NC algorithm, a single point is selected followed by arrangement of all other points (ranging from the nearest point to the farthest point). Later, Gabriel's graph is used to extract all direct and indirect relationships for all data points, and their density is measured in a single set. In this process, a new neighbourhood group is formed when the density inside the set or among the points decreases (decrease point is also termed as break point). In the newly-emerged inter-set neighbourhood groups, the neighbourhood formed in the intersection of break points and the neighbourhood group of a given point is carefully examined. In case any intersection is formed among the mentioned neighbourhood points, they will be considered as neighbours. If the mentioned condition is not fulfilled, and there is no intersection, independent and separate neighbourhood groups will be formed. In the last step of constructing algorithm, in order to identify the outlier data, the focus is on examining the mutual relationships among the set of nearest neighbours. In this way, it is possible to extract and identify the outlier data. This algorithm suffers from being too complex; therefore, it is especially useful in examining the data sets in small databases.

In the field of computational geometry and geometric graph theory, $\beta$-skeleton refers to the skeleton of the graph which is determined according to some Euclidean geometric points. In this regard, the value of $\beta$ should be determined first. By using $\beta$ value for point $X_i$ and $X_q$ in Eq. (4), θ is determined, and by definition, for the angles of $90^0$ and higher, neighbourhood is defined as the intersection region the two points of the subtended arc circles. For the angles of less than $90^0$, neighbourhood will be defined as the union of two subtended arc circles. In the case of $90^0$degree angles, the obtained angle works like Gabriel's algorithm. If $\beta$ value in $\beta$-skeleton algorithm is more than *1*, the nearest neighbourhood is defined as graph neighbouring area. In this algorithm, there is a need to regulate $\beta$ parameter (Langerman 2009, Toussaint 1988, Yang et al. 2016). Fig. 4 shows the performance of this algorithm with different β values.

$$\theta = \begin{cases} \sin^{-1}\dfrac{1}{\beta} & if\ 1 \le \beta \\ \pi - \sin^{-1}\beta & if\ \beta \le 1 \end{cases} \tag{4}$$

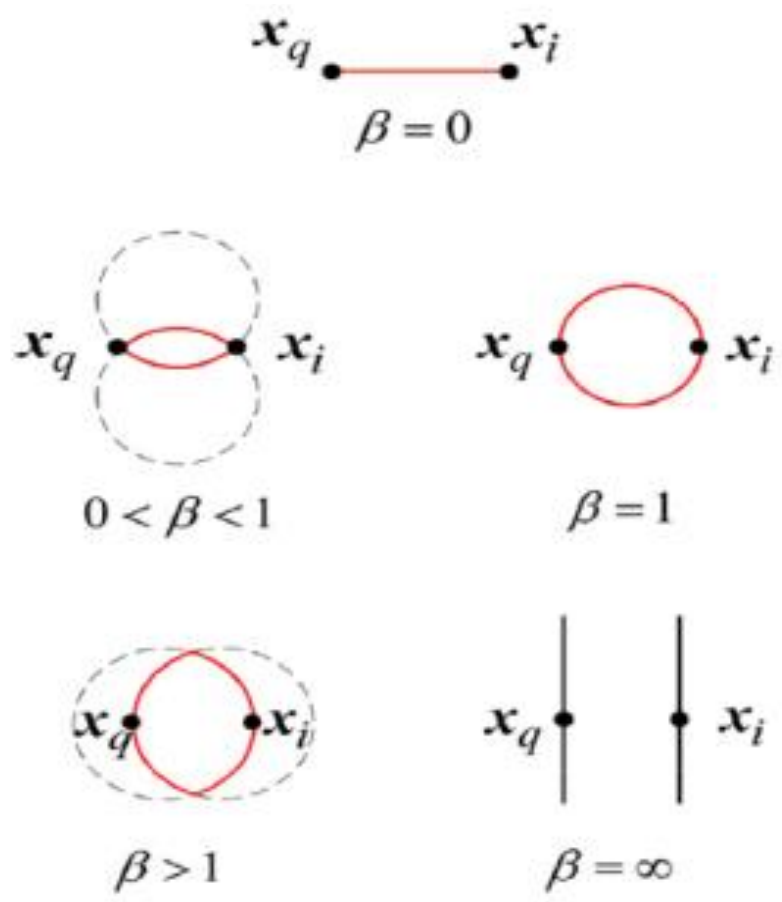

**Fig.4.** Beta skeleton (Yang et al. 2016).

In Fig. 5, the relative neighbourhood graph neighbours (RNGN) is displayed. In RNGN, *a* and *b* are determined as the centers of two circles, and the distance between *a*, *b* are the radius of the circles, and if there are no other points in-between, these two points are considered as neighbor points (Yang et al. 2016 ). However, the problem with this algorithm and Gabriel's graph is having fixed neighbourhood area, which prevents optimal efficiency in finding neighbourhood points. Gabriel's graph algorithms and RNG do not need to regulate their parameters.

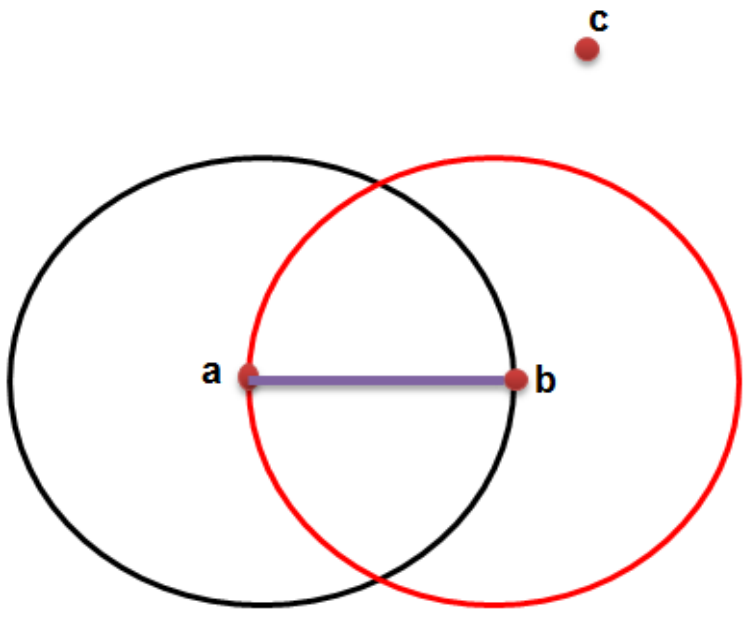

**Fig.5.** RNGN.

Another algorithm called Angle-based Neighbourhood Graph (ANGN) finds neighbourhood of data points according to the angle between the points (Yang et al. 2016 ). This algorithm uses geometric relationships and by defining the parameter, it extracts the angle between direct and indirect geometric points. Angle θ between $x_i$ and $x_q$ is determined on the circle of subtended arc, and if a three-point angle of $x_i, x_j, x_q$ is smaller than the two-point angle of $x_i, x_q$ these two points will have a direct interrelationship (using Eq. (5)). The advantage is its size of neighbourhood which can be adjusted according to the angle parameter, its efficiency in determining optimal precision in the data point neighbourhood. Fig. 6 and Fig. 7 show angle neighbourhoods with different angle parameters. The disadvantages of this algorithm include the regulation of the angle parameter and its complicated calculations.

$$\angle x_q x_j x_i = \arccos \frac{(x_j - x_i) - (x_q - x_i)^T}{\|x_j - x_i\| \cdot \|x_j - x_q\|} < \alpha, \forall_j \neq i, j = 1, 2, ..., N \quad (5)$$

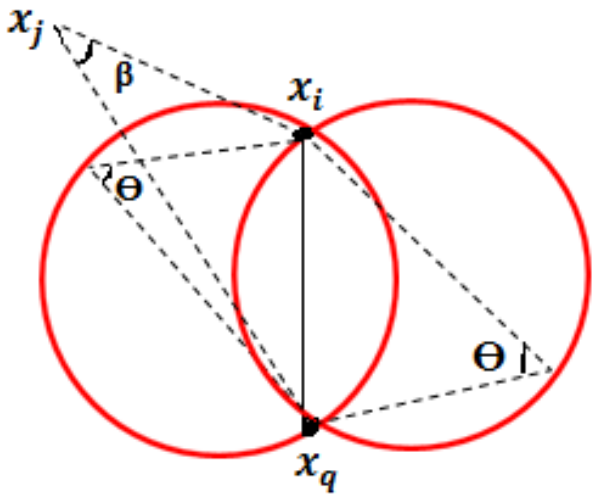

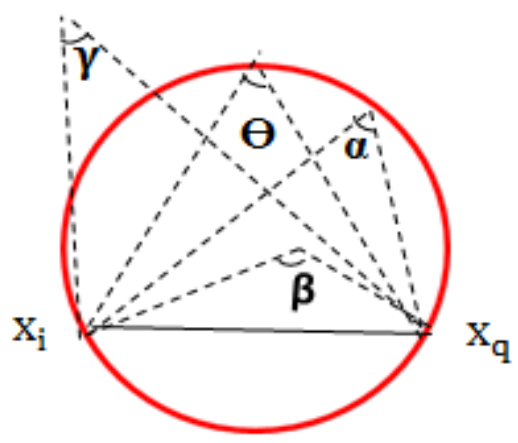

**Fig.6**. Example of ANGN angles (Yang et al. 2016). **Fig.7**. Chord and related angles (Yang et al. 2016).

If two points on the plate share a Voronoi edge, they are neighbours. This type of neighbourhood is what Delaunay Triangulation (DT) is based on, where each triangle edge represents a topology of the neighbourhood of the two points. When a neighbourhood is made with a Voronoi diagram, the space is divided into a number of regions. Each data point is allocated a region called as a Voronoi cell. For each data point from the set of points in the defined region, all the edges in the region are closer to the point that has generated the region. One of the applications of Voronoi diagram in DT is in the field of point clustering. In computational geometry, the DT for a finite set of points P on a plane R is a triangulation DT (P) such that none of the triangulation points of P is located inside any of the circumscribed circles of triangulation DT (P) (Liu et al. 2008). Based on Fig 8(a) and (b), a finite set of points in a Euclidean space is assumed, and a region for each point is postulated. The regions contain the points whose distance from each point inside the region from a specific point is less than or equal to the distance from the other edges of the concerned points. Two points would be located in a single cluster if they are in the vicinity of the convex hull, and share at least one edge. DT needs *O (nlogn)* time complexity for $d \leq 3$ and $O(\frac{n^{\lfloor \frac{d}{2} \rfloor}}{\lfloor \frac{d}{2} \rfloor !})$ for $d \geq 3$.

In algorithm AMOEBA, the standard deviation and mean of edge lengths are used to detect clusters based on edges DT (Estivill-Castro and Lee 2000). In this algorithm, featuring the edges in Delaunay Triangulation, local effect and global effect are detected. AMOEBA is a parameter-free and graph-base algorithm. AMOE has one disadvantage that is designed for 2-dimensional data sets and the worst case will take at least *O (nlogn)* time complexity.

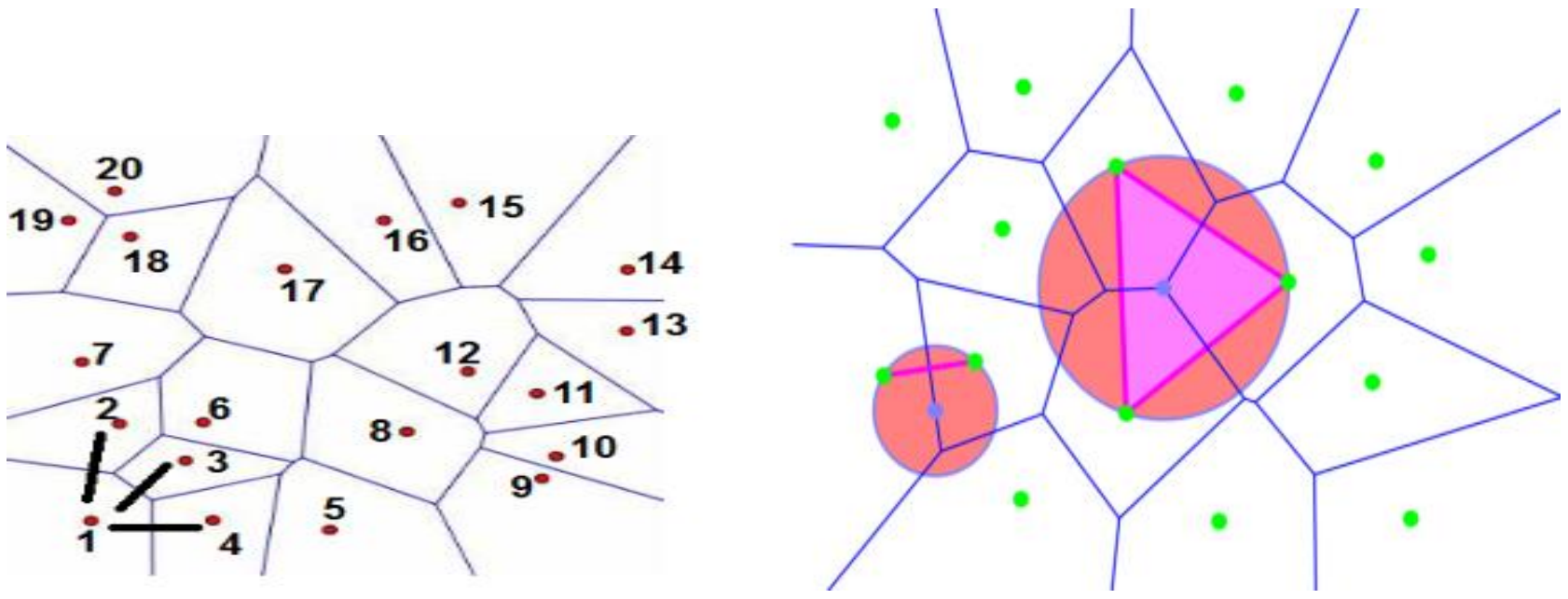

**(a)** **(b)**

**Fig. 8**. (a) Example of Voronoi diagram (contains the *NN*-boundary), (b) Delaunay triangulation .

At first, high density points labelled as target points are located by Neighbourhood Construction in Apollonius Region (NCAR). NCAR initially locates the most distant point (farthest point) for each given target point and determines both the center and radius of Apollonius circles accordingly, which finally leads to forming primary Apollonius circles (Pourbahrami et al. 2018). The problem lies in the fact that NCAR algorithm does not work accurately enough with large scale data. Therefore, to alleviate the problem, the criteria of density and internal relationships inside the circles were used as the basis. Therefore, in each single point near the target points, it is possible to draw Apollonius circle and define safe neighbourhood zone. In this algorithm, by the use of Apollonius method, it is feasible to find the accurate neighbourhood area among the points. For small scale data, this method is more effective than the other methods. Moreover, this method can locate and identify the outlier data very easily. The shape and distance among the points determine the structure of NCAR algorithm and are useful in offering more accurate and new definition for neighbourhood. Besides, neighbourhood is determined more accurately by locating and identifying interconnections among the points. This means that by examining the density of the points in Apollonius circle, accuracy is enhanced. NCAR algorithm does not need to utilize data knowledge and is used in clustering process. Fig. 9 shows NCAR algorithm by Apollonius circles.

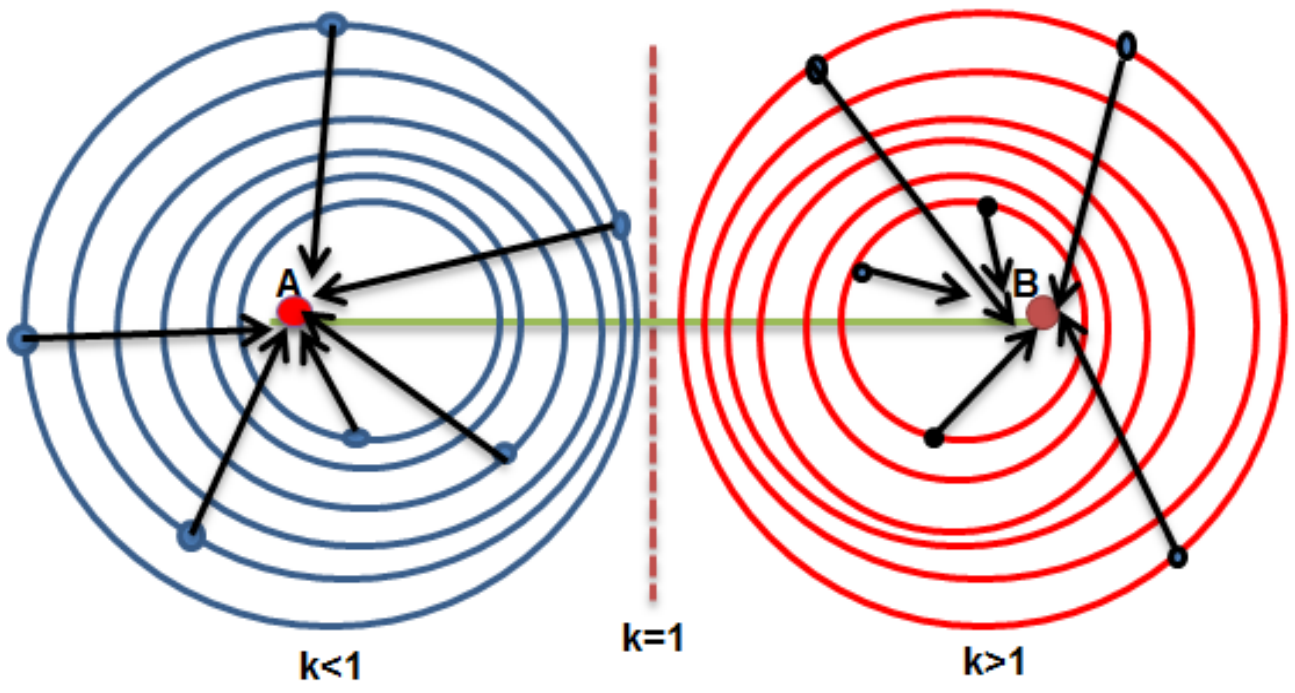

**Fig. 9.** Apollonius circles $C_{AB}$ (Pourbahrami et al. 2018).

## 4. Neighbourhood Construction: Artificial Neural Network-based Approaches

To sum up, the work done regarding data categorizing can also be shown in these groups. Machine learning algorithm of the groups is among the algorithms suggested in these papers (Wang et al. 2016, Luo et al. 2016, Huang et al. 2015, Peng et al. 2016, Esmin et al. 2015, Rana et al. 2011). Most of these algorithms use neural networks (Shen and Hasegawa 2008, Singh et al. 2016) and support vector machine for classification ( Güney and Atasoy 2012, Pan et al. 2015). Support Vector Machine (SVM) is an algorithm for finding data point neighbourhood with a supervisor. It is one of relatively new algorithms which have recently shown greater efficiency in finding data point neighbourhood including perceptron neural networks (Varma et al. 2016). Chatterjee and et al. employed a similarity graph neighbourhood by determining displacement for each element in the training feature subspace and mapped the input data set and finally trained a classifier on the transferred data point (Chatterjee and Raghavan 2012). They proposed a data transformation algorithm to optimize the accuracy of SVM.

Self-Organizing Map (SOM) is an algorithm which uses competitive learning algorithm for training and is developed according to specific features of human brain. The main difference between SOM training algorithm and other vector measurement algorithms is the fact that besides a single communication weight, they highly match each other, and the weights of the neighbouring winning cells are updated. Close observations in the input area activate the two neighbouring units of the map (Oja et al. 2003, Hajjar and Hamdan 2013). Training phase continues until the weight vectors reaches a stable state and do not change anymore. After the training phase, in the mapping phase, it would be possible to rank each vector of the input vector automatically (Roigé et al. 2016). Since SOM has high visualizability, it dominates the approaches based on artificial neural (Awad 2010, Vesanto and Alhoniemi 2000). The projects involving SOM involve a set of data highly dimensional and located on common vectors of a low dimensional grid structure. There is a grid for each neuron. The network involving SOM is arranged in a way that similar points are assigned for neuron .Two adjacent neurons on the grid lead to close data points in the main data. Therefore, SOM is useful in visualizing and finding neighbourhood points in the data sets. The main purpose of this method is reducing data dimensions to one or two. This method is also called as Self-Organized Neural Network; it is, in fact, a subset of artificial neural network. Advantages: determining the optimal classes in big data sets. Disadvantages: One of the main disadvantages of model-based methods is their sensitivity to selection of primary parameters of the model and the number of clusters. Moreover, they are not able to identify the clusters with different shapes.

## 5. Neighbourhood Construction: Density-based and Graph-based Approaches

Optimum Path Forest ( $OPF_{knn}$ ) uses graphs to find data point neighbourhood. Initially, by the use of the distance between the nodes, $OPF_{knn}$ forms complete graph (Papa et al. 2017). Then, the spanning tree of the points is obtained, and dense nodes are identified. Later, the cost of each node or the cost of distance between two nodes is identified (Zero cost for dense node and infinite cost for normal nodes). Data point neighbourhood can be located in the training step of the algorithm through either minimization or maximization of the costs, and the groups can, subsequently, identify the conductivity of testing node based on minimum or maximum cost rate in the groups. Fig. 10 (a)-(d) shows $OPF_{knn}$ classification process.

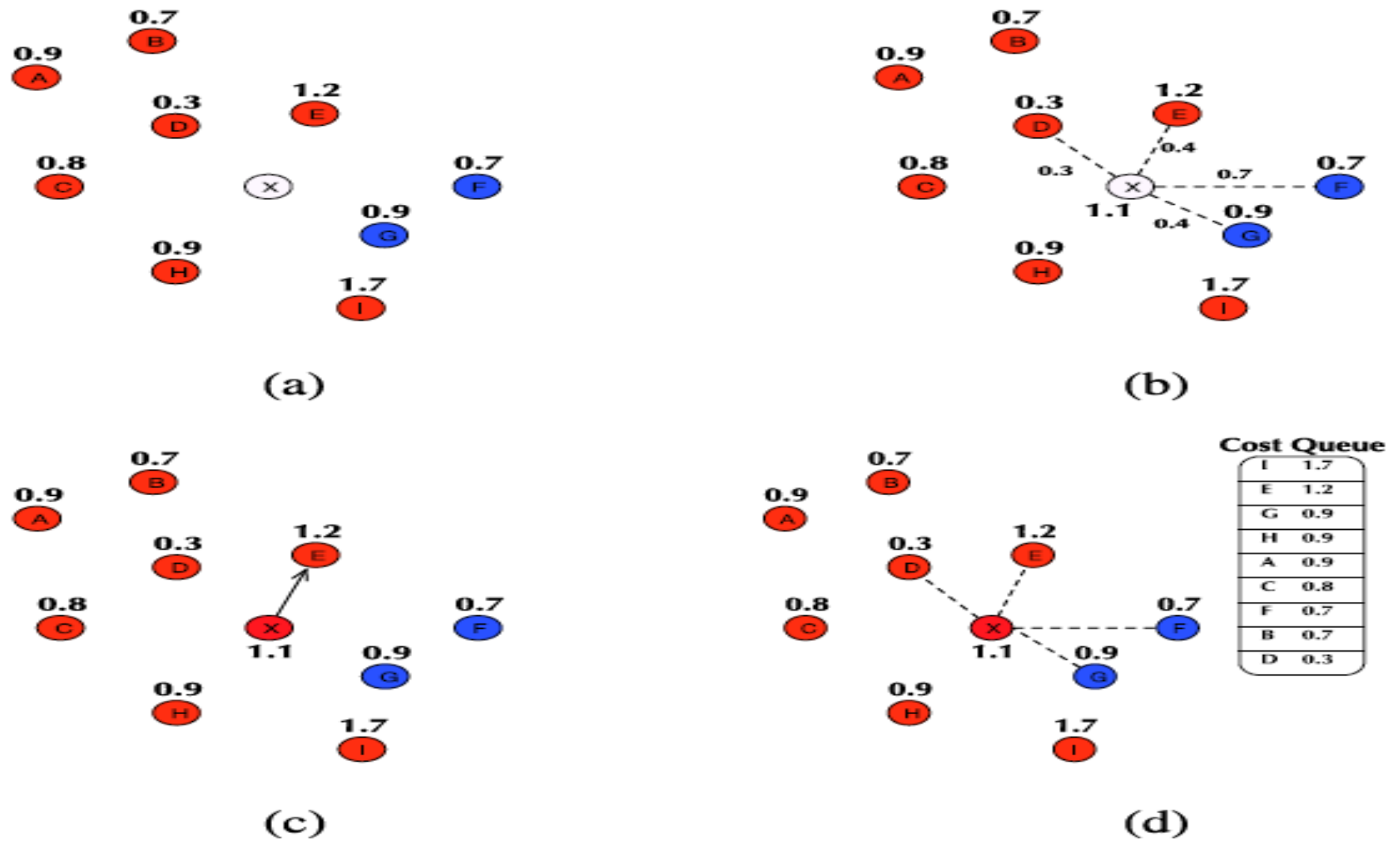

**Fig. 10**. $OPF_{knn}$ classification process: (a) training samples from classes “red” and “blue”, and one testing sample (“white”) to be classified, (b) testing sample is connected to its 4-nearest neighbours, for further computation of its density, (c) testing sample is conquered by sample ‘E’, and then labeled as belonging to class “red”, and (d) cost queue is used to speed up the classification step (Papa et al. 2017).

Since ε-neighbourhood cannot find clusters with different density levels effectively and cannot deal with several outliers that satisfy the density requirements, the Karypis is proposed to remedy the ε-neighbourhood deficiencies (Karypis et al. 1999). It first uses a KNN graph to generate the sub-clusters and then determines the similarity between sub-clusters by looking at their Relative Interconnectivity (RI) and Relative Closeness (RC); it lastly merges the sub-clusters with high similarity. Other works using KNN graph are Graph-based MultiPrototype Competitive Learning (GMPCL) (Wang et al. 2012) and Shared Nearest Neighbours clustering (SNN) (Ertöz et la 2003) (Xu et al. 1998, Franti et al. 2006, Chen et al. 2009). GMPCL relies on KNN graph to initialize large clusters and uses GMPCL to refine the large clustering. SNN utilizes the shared-nearest-neighbor graph to discover the core points and generate clusters based on these core points. Hence, there are many clustering methods based on Mutual K-NN (MKNN) graph (Huang et al. 2014, Abbas and Shoukry 2012, Brito et al. 1997, Hu and Bhatnagar 2012, Sardana and Bhatnagar 2014). Clustering using MUtual nearest NEighbours (CMUNE) (Abbas and Shoukry 2012) utilizes MKNN graph to calculate the density of each point and chooses the high-density data points as the seeds from which clusters may grow up. Clustering an algorithm for Arbitrary ShaPed clusters (CLASP) (Huang et al. 2014) is a three-phase clustering method. During the first and second phases, CLASP shrinks the original data sets and uses a position adjusting method to make the clusters structures clearer and more distinct. During the third phase, it uses an MKNN graph to discover the arbitrarily shaped clusters. Since the above mentioned graph-based methods just use a single type of nearest neighbor graph, they can only discover either connectivity information or density information contained in the data sets. However, the proposed Hybrid K-NN (HKNN) graph which combines the advantages of MKNN graph and KNN graph can discover the density and connectivity information simultaneously. Moreover, the CHKNN can process the clustering problems on more complex and noisy nonlinear data sets than other graph-based methods.

Most of the parameters of clustering methods are difficult to be determined in practice. In order to address such problems, a large number of internal validity indicators are proposed for choosing the optimal parameters (Davies and Bouldin 1979, Rousseeuw 1987, Bandyopadhyay and Saha 2013, Rezaee 2010, Schaeffer 2007). The widely used internal validity indicators such as Davies-Bouldin Index (DBI) (Davies and Bouldin 1979) and Silhouette Coefficient (Sil) (Rousseeuw 1987) are based on the correlation of inter-clusters and intra-clusters. Both of these two correlations just consider the distance parameter ignoring the connectivity information among data points, hence the correlation of intra-clusters in a circular cluster is higher than that of annular clusters. Therefore, these two indicators are unable to evaluate the validity of clustering results on the nonlinear data sets like concentric ring data sets correctly. Another deficiency of most of the existing internal indicators is that they do not have any input parameters. In general, there is no golden standard for clustering results, and the optimal clustering results are different based on different situations and different criteria. Consequently, a certain criterion is necessary when the validity of clustering result needs to be evaluated. To remedy for the above mentioned deficiencies, the Qin propose an internal validity indicator termed K-Nearest Neighbour Index (KNNI) is used; it first constructs a KNN graph using an input parameter M, it then utilizes the connectivity information among this KNN graph to evaluate the validity of clustering results. The input parameter M represents the number of nearest neighbours of each point, and it controls the number of edges in a KNN graph (Qin et al. 2018). In HKNN, a novel data model termed Hybrid K-NN (HKNN) graph, which combines the advantages of mutual K-NN graph and K-NN graph, is proposed to represent the nonlinear data sets. Moreover, a Clustering method according to the HKNN graph (CHKNN) is proposed. The CHKNN first generates several tight and small sub-clusters, and then merges these sub-clusters by exploiting the connectivity among them. Fig. 11 shows CHKNN clustering process.

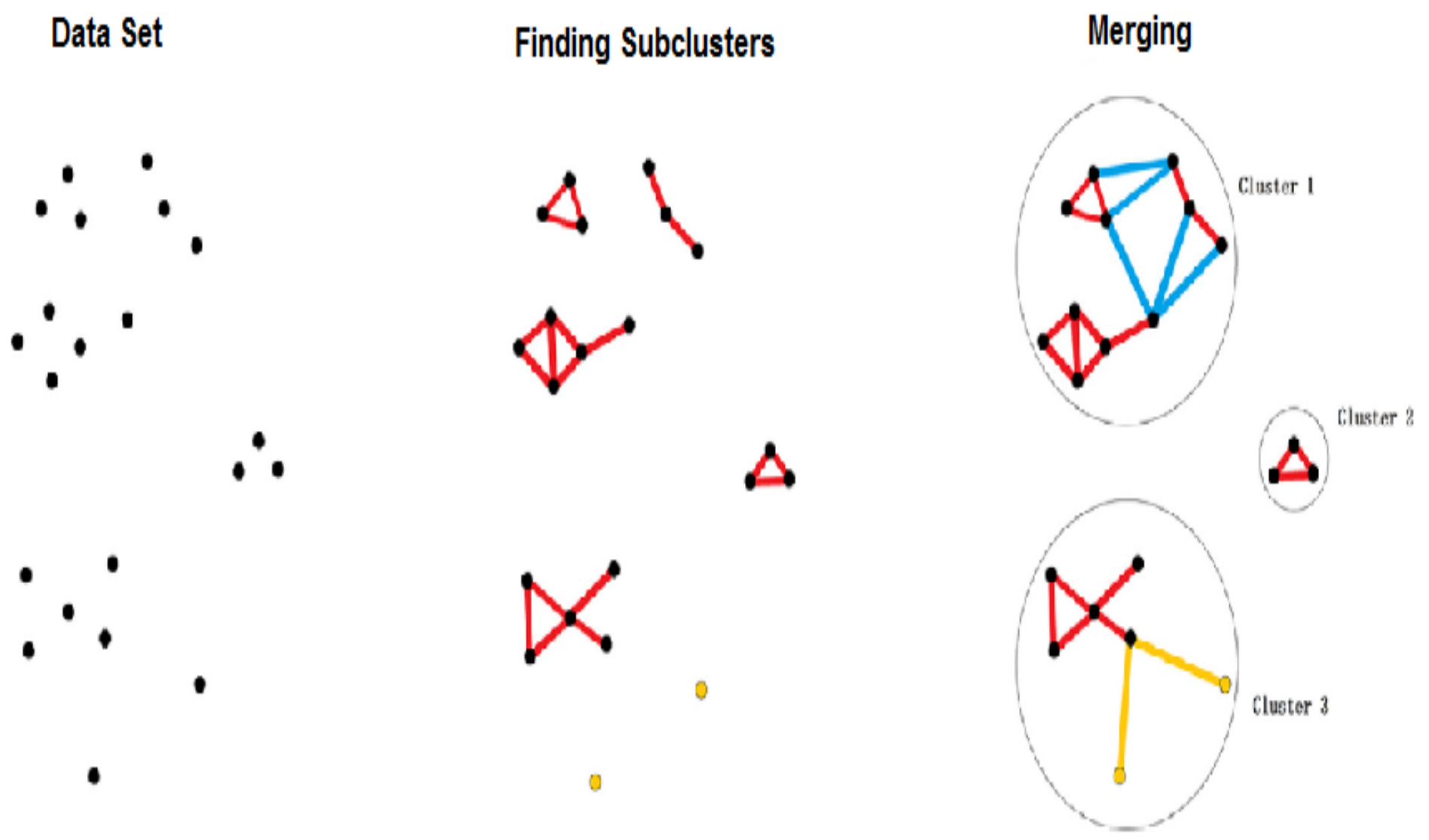

**Fig. 11**. CHKNN clustering process. In the schematic of finding sub-clusters phase, the red lines connect the points in the same bi-sub-clusters , while the yellow points represent the isolated points. In the schematic of merging phase, the blue lines connect the bridge points from one bi-sub-cluster to another bi-sub-cluster, while the yellow lines show that the isolated points are assigned to their nearest main cluster (Qin et al. 2018).

The number of neighbourhood is determined based on K-associated Optimal Graph. It starts with 2, which enhances the frequency of *ks* in each step. The frequency of *ks* is obtained from the nodes and it, subsequently, estimates the rate of purity (estimated according to input and output nodes and their frequency) (Mohammadi et al. 2015, Nettleton 2013). In every step, with an increase in *K*, an integration among the groups happens (i.e., purity is calculated again). In case, there is an increase in purity, and groups are integrated suitably. If there is no increase in the value of *K*, no integration or change will happen.

At the top left corner of Fig. 12, a part of an optimal graph is shown with two groups of $\mathrm{C}_1$ and $\mathrm{C}_2$. Group $\mathrm{C}_1$ is generated according to three samples, with *K* = 2. Group $\mathrm{C}_2$ has 5 samples, and the *K* value is 2. The other samples which are marked in green are related to other classes. The purity value for $\mathrm{C}_1$ and $\mathrm{C}_2$ are $\Phi_1$ and $\Phi_2$, respectively. At the top right corner of Fig. 12, a *K* associated graph with *K* = 3 is shown. As shown in this figure, group $\mathrm{C}_1$ and $\mathrm{C}_2$ tend to integrate together in $\mathrm{G}^{(\mathrm{K}=3)}$ when the *K* value is 3. If $\mathrm{C}_1$ and $\mathrm{C}_2$ are integrated, the purity of the final group ( $\mathrm{C}_1^{(\mathrm{K}=3)}$) is *b*. If the purity of the new generated group is higher than the purity of $\mathrm{C}_1^{(\mathrm{opt})}$ and $\mathrm{C}_2^{(\mathrm{opt})}$, the integrating process will be completed, and the $\mathrm{G}^{(\mathrm{opt})}$ will be updated. Otherwise, $\mathrm{C}_1$ and $\mathrm{C}_2$ remain the same in $\mathrm{G}^{(\mathrm{opt})}$. In KAOG algorithm, the input is the data set, and the output is the optimal graph consisting of some groups ( $\mathrm{C}_\alpha^{(\mathrm{opt})}$). Each group has a subgraph drawn according to a different value of *K*, which means that different groups may have different *K* values.

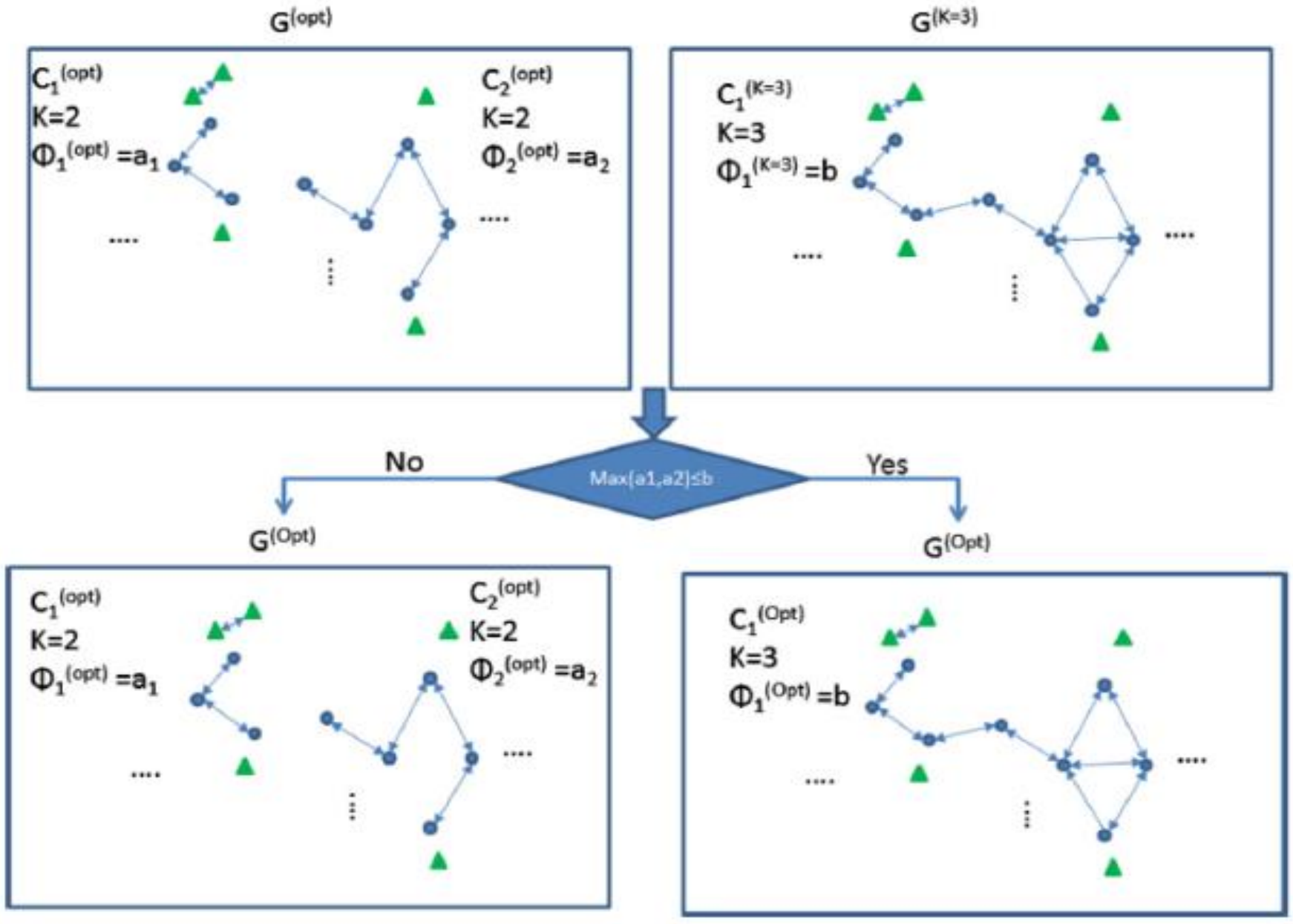

**Fig.12**. The integrating process in KAOG algorithm (The groups which are allowed get integrated in KAOG algorithm) (Mohammadi et al. 2015).

The new group can be measured using a new algorithm. The previous groups can be integrated into each other only if the new group has a purity level which is more than or the same as each newly merged group. If the mentioned conditions are not met, there is no possibility that the process of integration will continue. There are many situations in which the specified conditions are not fulfilled. However, the mentioned case is very similar to a real state of data sets. Fig. 13 indicates that the class samples have a common class and are built accordingly $(c_1,c_2,c_3,c_4)$ . In this figure, *K* has a value of 3. Since all of the samples within this group are linked to each other, some of the samples seem to connect themselves to the groups from the other class leading to purity value of 0.88. This value indicates good interconnection among most of the groups. The mentioned groups can be easily connected to each other to compose a bigger group. Since $c_1$ has a purity measure of *1*, it cannot integrate with the groups whose purity measure is less than one. Therefore, it is not possible to merge the groups in this algorithm .The experiments with real data sets show that the algorithm does not allow merging of the groups. In small data sets such as IRIS, most of the groups are small and consist of two or three samples. This means that if the purity measure for the newly merged group is higher than the average of purity measures for all the groups in the merging process, the groups gain the possibility to be merged. They change the merging condition and are calculated with Eq. (6) as follows:

$$if\left(\varnothing_{\beta}^{(K)} \geq \frac{1}{N}\sum \varnothing_{\alpha}^{(opt)}\right) for\ all\ C_{\alpha}^{(opt)} \subseteq C_{\beta}^{(k)} \tag{6}$$

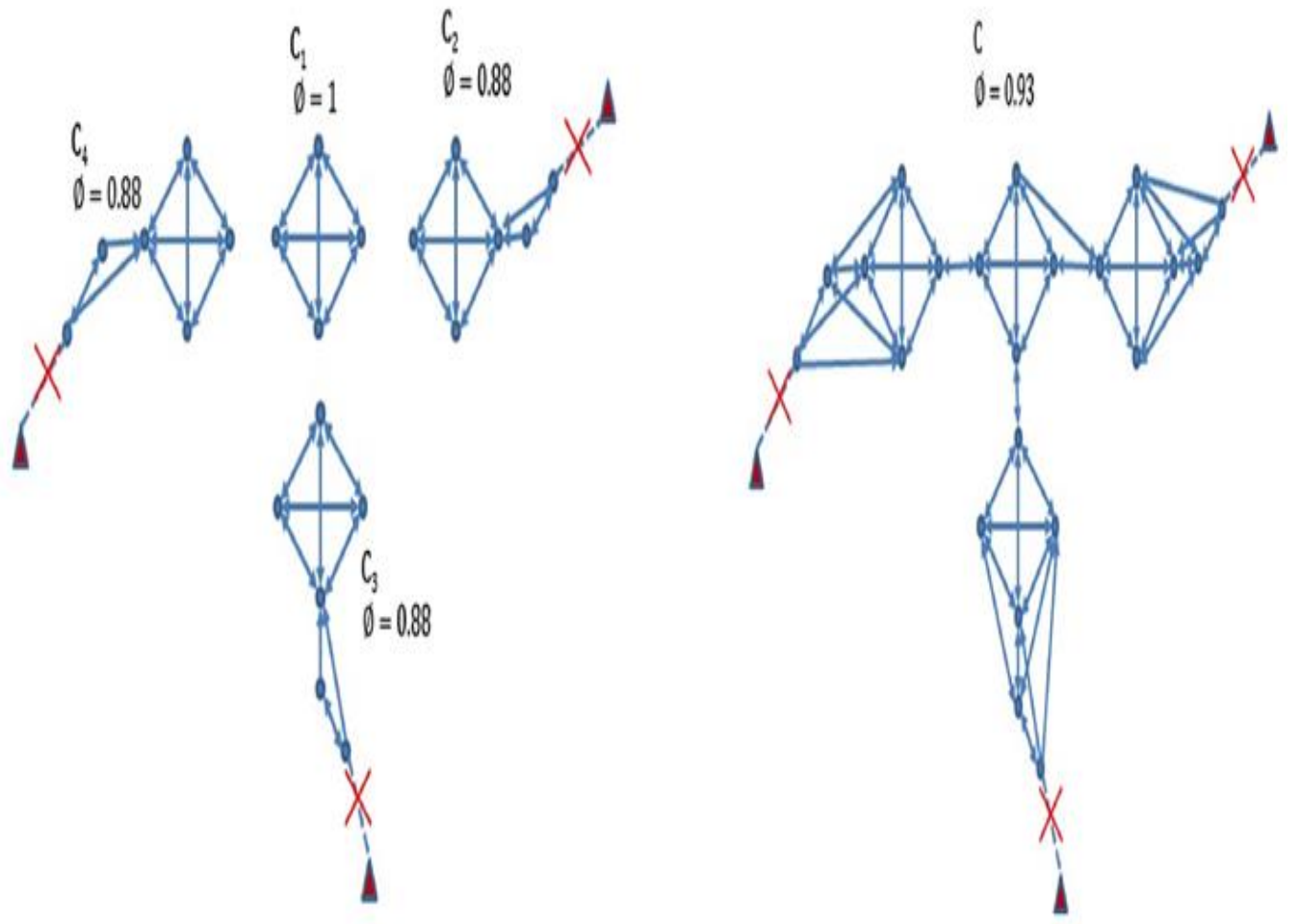

**Fig. 13**. The groups which are not allowed to be integrated in KAOG algorithm (Mohammadi et al. 2015).

MKAOG algorithm integrates the four groups (in Fig. 13) as shown in Fig. 14. The purity of the final group is 0.93, which is a relatively high value. The MKAOG algorithm is less sensitive to the input noise after small groups have integrated.

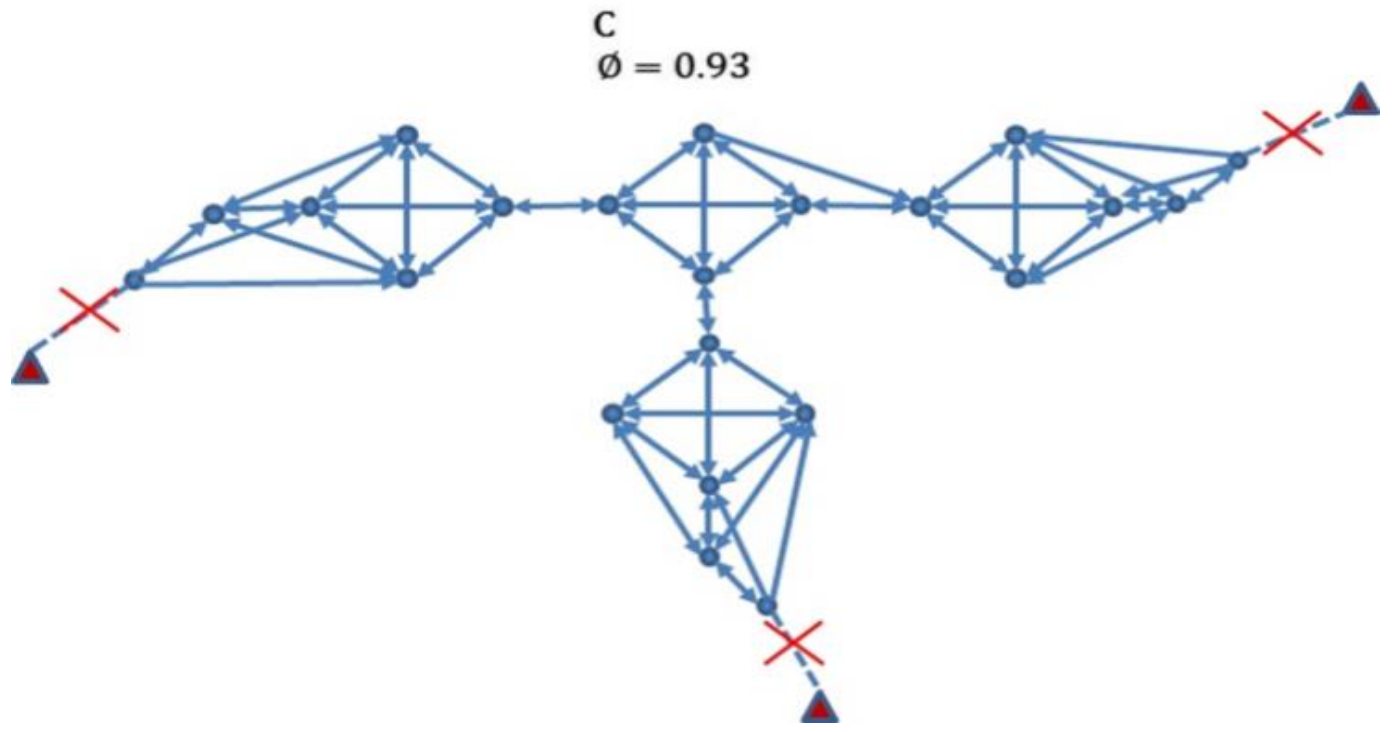

**Fig. 14**. Integrating groups in the modified KAOG (MKAOG) algorithm (Mohammadi et al. 2015).

In their papers (Yu and Kim 2016 a, b), Yu and Kim used mutual *k*-nearest neighbourhood for locating clusters according to the intensity of the points and the criteria. The algorithm labeled as Density-based Noisy Graph Partition (DENGP) has 5 steps. The five main steps of algorithm are shown in Fig. 15 (a)-(e)

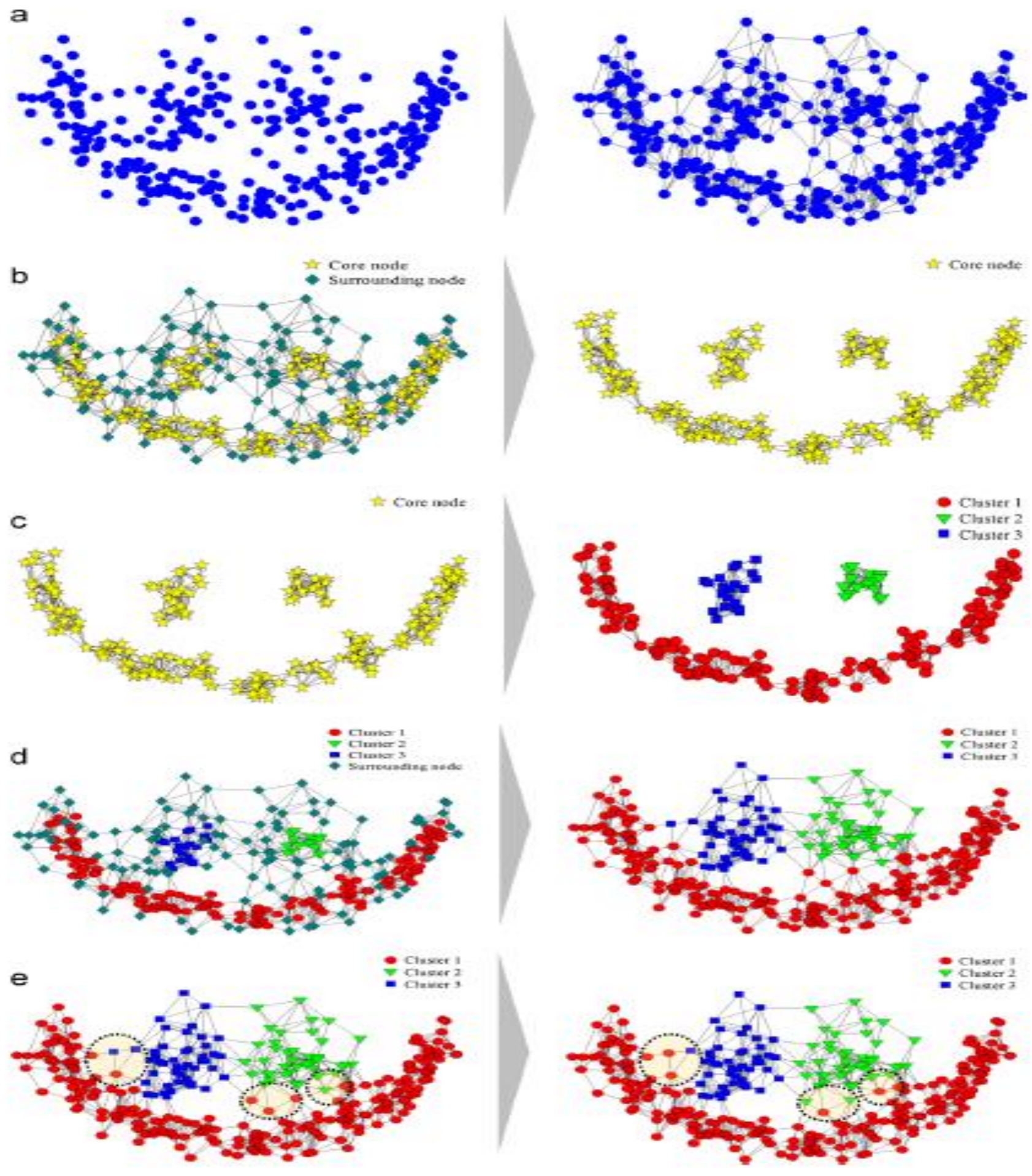

**Fig. 15**. Graphical illustration for the proposed clustering algorithm with example data set: (a) Constructing mutual *k*-nearest neighbour graph from the data set, (b) Classifying the nodes using density coefficient, (c) Partitioning the core nodes into three subgroups, (d) Assigning the surrounding nodes, and (e)Reassigning incorrectly assigned nodes (Yu and Kim 2016 a)

The first step: The data with graph structure is illustrated with the *k* nearest to the mutual neighbourhood calculated using the following formula for node $x_i$ .

$$e_{ij} = \begin{cases} 1 & if\ x_i \in K(j) and\ x_j \in K(i) \\ 0 & otherwise \end{cases} \tag{7}$$

The second step: For each node the density coefficient is calculated for determining the local areas with high and low intensity in the structure of the graph. The following formula calculates the density coefficients for node *i*.

$$d_i = \sum_{j \in N(i)} W_{ij} + \sum_{j,k \in N(i), j \prec k} W_{jk} \tag{8}$$

The third step: The core nodes as well as neighbouring nodes from the second step are detected, and the neighbouring nodes are removed (based on a given threshold level) to categorize the core nodes into original subgroups.

The fourth step: After forming core node clusters, the nodes neighbouring the core nodes are re-added using modularity weighting and voting plan.

$$W_{ij} = \exp\left( -\frac{d\left(x_i, x_j\right)^2}{d\left(x_i, x_i^k\right) d\left(x_j, x_j^k\right)} \right) \tag{9}$$

The fifth step: Transfer of the neighbouring nodes to the clusters is examined, and if the transfer is not accurate, they will be transferred again according to the weighting criteria.

Central Diagram Algorithm (CDA) is another type of algorithm that indicates direct/indirect relationships among the points by the use of visualization. In order to determine neighbourhood, CDA provides an even distribution of nodes over the circles with central nodes and defines various nodes shown in communication layers (Park and Basole 2016). For each circle center, there are two basic nodes in the database. The centers shown in Fig. 16 (a)-(d) determine the relationship of other visualized nodes. Although direct or close relationships are shown with inner layers, rather distant relationships are indicated with outer layers (indirect connection or lack of connection). Finally, direct/indirect relationships are identified (Crnovrsanin et al. 2014, Vehlow et al. 2015). In Fig. 17 the taxonomy of neighbourhood construction approaches is displayed.

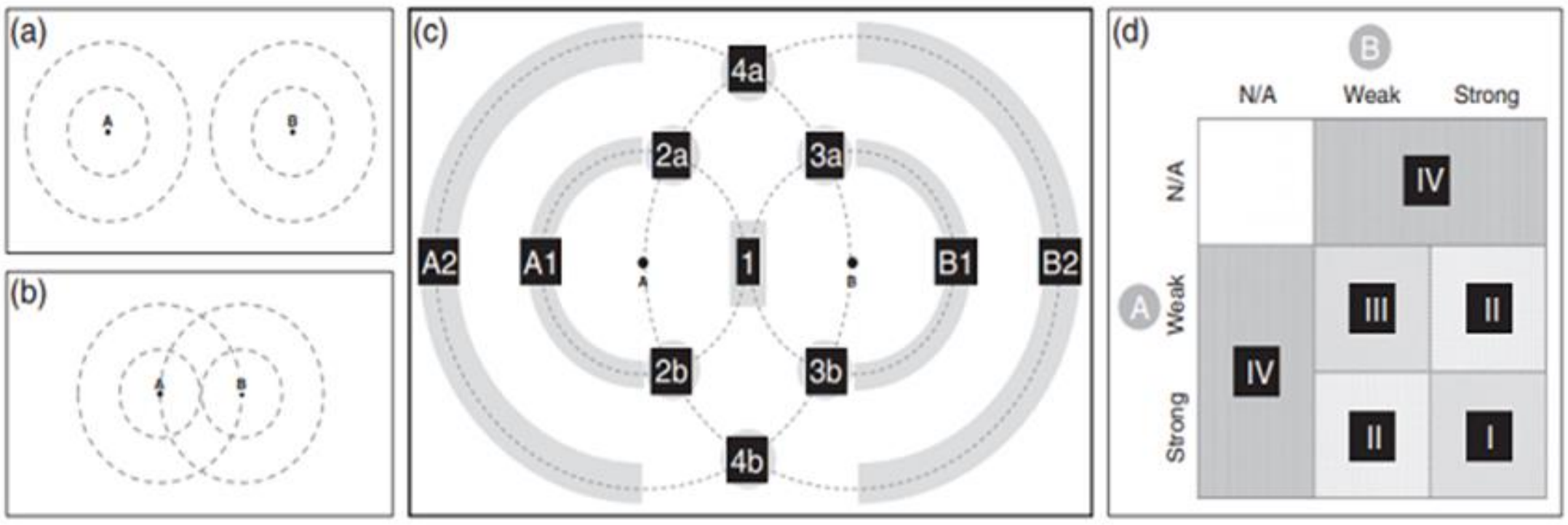

**Fig. 16.** (a, b) The bicentric layout builds on the concentric network visualization to provide an efficient representation of two focal points as well as their shared direct and indirect points while organized by tier. (c) This schematic representation provides an overview of point placements in the bicentric layout. (d) Classification of relationship types identifies four combinations of tie strength (Park and Basole 2016).

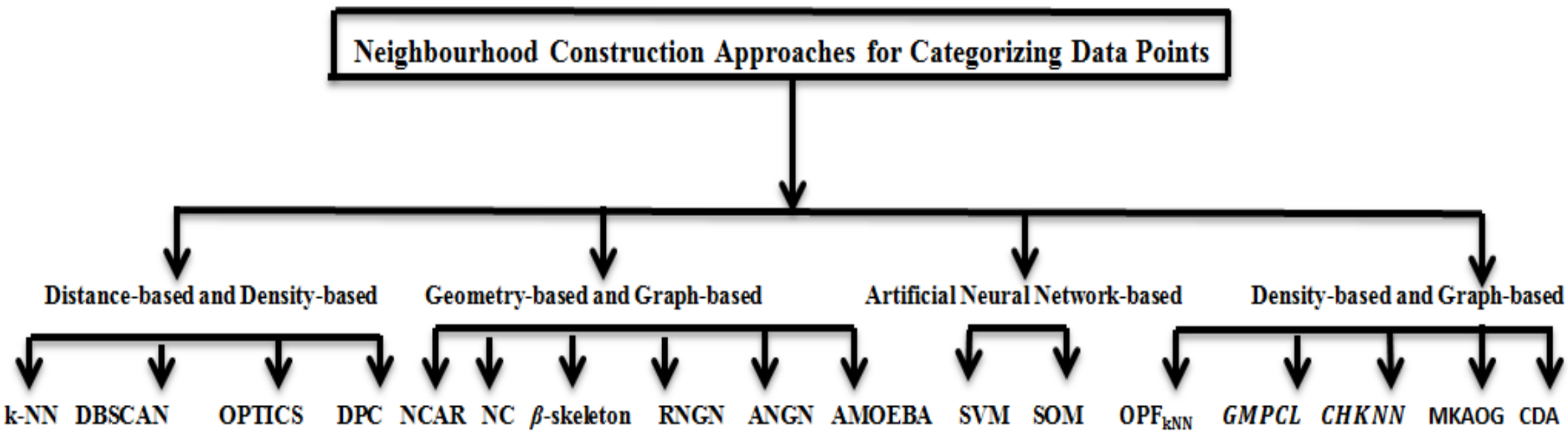

**Fig. 17**. The Taxonomy of neighbourhood construction approaches.

## 6. The Challenges and Directions for Neighbourhood Construction Approaches

Table 2 compares the mentioned neighbourhood construction algorithms in previous sections based on the worst case run time complexity. In Table 2, $n$ is the cardinality of the observations set, and $d$ refers to the dimension of each sample. Comparison between the algorithms detailed in the paper demonstrates that the $k$-NN algorithm is simple and versatile and has low complexity, but it depends upon how parameter $k$ is set. As observed in Tables 2, NC algorithm, using the Gabriel graph for constructing neighbourhoods between points has high complexity, since it examines the neighbourhood of a point with respect to all the points in the data set, and is therefore inappropriate for large data sets. The AMOEBA and OPTICS algorithms were among the neighbourhood construction algorithms for clustering, and have low complexity; it is, therefore, a proper choice for large data sets. The $\beta$-skeleton algorithm, which determined the neighbourhood based on parameter $\beta$, has high complexity, since it examines all the points for determining the neighbourhood. One of its advantages, however, is that the neighbourhood areas can change with *β* parameter, which can improve neighbourhood definition through right definition of *β* parameter. The RNGN algorithm has $o(n^3)$ complexity, which offers improper times for neighbourhood computation in large data sets. Like the SVM and ANGN algorithms, this algorithm has high complexity, since it examines the neighbourhood of a point with respect to all the points in the data set. SOM and $OPF_{knn}$ have $o(n^2)$ complexity that is appropriate for medium-sized data sets. KAOG and MKAOG algorithms, used for classifying points based on neighbourhood specification, show high accuracy over different data sets. To reduce the high complexity and computation in these algorithms, one can pursue geometric and mathematical algorithms that are not based on parameter specification, and where neighbourhood areas are not fixed. Algorithms that do not examine all the relations between the points and operate locally will be appropriate algorithms for the reduction of complexity. Finally, NCAR algorithm was among the neighbourhood construction algorithms for clustering, and has low complexity.

**Table 2. Computational complexity of neighbourhood construction approaches**.

| Neighbourhood approaches | Time complexity | Reference | Type of categorizing |
|---|---|---|---|
| *k*-NN | $o(nkd)$ / $o(n^2)$ | Von Luxburg 2007, Duda et al. 2012 | Clustering |
| ε-neighbourhood | $o(n^2)$ | Pedrycz 2010 | Clustering |
| OPTICS | $o(n\log n)$ | Ankerst et al. 1999, Januzaj et al. 2004, Kriegel and Pfeifle 2005 | Clustering |
| NC | $o(n^3)$ | İnkaya et al. 2015 | Clustering |
| GGN | $o(d\,n^3)$ | İnkaya 2015a | Classifying |
| AMOEBA | $o(n\log n)$ | Estivill-Castro and Lee 2000 | Clustering |
| *β*-skeleton | $o(n^3)$ | Langerman 2009, Toussaint 1988, Yang et al. 2016 | Classifying |
| RNGN | $o(n^3)$ | Yang et al. 2016 | Classifying |
| ANGN | $o(n^3)$ | Yang et al. 2016 | Classifying |
| SVM | $o(n^3)$ | Güney and Atasoy 2012, Pan et al. 2015 | Classifying |
| SOM | $o(n^2)$ | Oja et al. 2003,Roigé et al. 2016 | Clustering |
| CHKNN | $o(kn^2)$ | Qin et al. 2018 | Clustering |
| $OPF_{knn}$ | $o(n^2)$ | Papa et al. 2017 | Classifying |
| KAOG | $o(n^3)$ | Mohammadi et al. 2015 | Classifying |
| MKAOG | $o(n^3)$ | Mohammadi et al. 2015 | Classifying |
| DPC-PCA | $o(n^3)$ | Du et al. 2016 | Clustering |
| NCAR | $o(n^2)$ | Pourbahrami et al. 2018 | Clustering |

Table 3 shows the accuracy of some well-known and state-of-the-art neighbourhood construction algorithms in clustering and classification in IRIS data set. The accuracy of *k*-NN over the data set is 81.06, which is relatively acceptable, but the shortcoming of the algorithm lies in its incorrect selection of *k*. The ε-neighbourhood algorithm exhibits the lowest clustering accuracy over the IRIS data set, but it has low complexity, and is appropriate for medium data sets. In this algorithm, incorrect selection of the ε parameter causes errors in making clusters. As observed in Table 3, NC algorithm, uses Gabriel graph for constructing neighbourhoods, and is, therefore, inappropriate for large data sets. However, it exhibits proper accuracy in clustering of small data sets, and its accuracy over the IRIS data set is 87.06. Algorithms using the Gabriel graph have fixed neighbourhood areas, which may make it difficult in future to optimally define neighbourhoods for them. In Table 3, their clustering accuracy of KAOG and MKAOG over the IRIS data set are 94.84 and 95.86, respectively. Table 4 shows the accuracy of some well-known and state-of-the-art neighbourhood construction algorithms in clustering and classification over the WINE data set.

**Table 3. Some categorizing results for the IRIS data set**

| Neighbourhood algorithms | Percentage of accuracy | Type of categorizing |
|---|---|---|
| *k*-NN | 81.06 | Clustering |
| ε-neighbourhood | 72.09 | Clustering |
| NC | 87.06 | Clustering |
| NCAR | 93.17 | Clustering |
| GGN | 71.06 | Clustering |
| RNGN | 92.81 | Classifying |
| ANGN | 93.33 | Classifying |
| KAOG | 94.84 | Classifying |
| MKAOG | 95.86 | Classifying |

**Table 4. Some categorizing results for the WINE data set**

| Neighbourhood algorithms | Percentage of accuracy | Type of categorizing |
|---|---|---|
| k-NN | 79.82 | Clustering |
| ε-neighbourhood | 70.49 | Clustering |
| NC | 88.75 | Clustering |
| NCAR | 90.18 | Clustering |
| GGN | 94.03 | Clustering |
| RNGN | 94.15 | Classifying |
| ANGN | 96.69 | Classifying |
| KAOG | 85.54 | Classifying |
| MKAOG | 89.86 | Classifying |

In Table 5, the advantages and disadvantages of the discussed categorizing approaches are displayed. Comparisons between neighbourhood construction algorithms in Table 5 indicate that most of the algorithms in base geometry have high computations except for AMOEBA and NCAR which have low computations, but they are only used for small scale data. In fact, AMOEBA and NCAR have the least complex algorithm among all neighbourhood construction algorithms used for the purpose of clustering. The geometric methods are able to identify the outlier data easily, and they use the geometric relationships among the data to increase the predictive power of neighbourhood. These methods do not require initial determination of the number of clusters, and algorithms such as NC and RNGN do not require determining a specific type of parameter. These algorithms use simpler approaches than the other distance-based and density-based identification methods, and DBSCAN, OPTIC, and KNN use fewer computations. Base density algorithms are able to locate the clusters having different shapes.

**Table 5. The advantages and disadvantages of the discussed categorizing approaches.**

| Neighbourhood Approaches | Advantages | Disadvantages |
|---|---|---|
| *k*-NN | • Simple approach<br>• Nonparametric<br>• Intuitive approach<br>• Robust to outliers detection<br>• Good at dealing with numeric and continues features<br>• Do not information | • Determining neighbourhood just according to the distance<br>• Sensitivity to irrelevant features<br>• Very difficult in managing data of mixed types<br>• Lack of theoretical basis<br>• Unclear on non-numerical/binary values |
| DBSCAN | • Discover clusters of arbitrary shapes<br>• Middle complexity<br>• Processing is fast<br>• Outliers detection<br>• No need to define number of clusters | • Cannot find clusters effectively with different density<br>• Cannot work with several outliers<br>• High computational cost when the data set is large<br>• Uncertainty of tuning the input parameters<br>• Not suitable for high-dimensional data<br>• There is not a unified and generally accepted approach to determine parameters |
| OPTICS | • Detecting meaningful clusters in data of varying density<br>• Low computational cost<br>• Robust to outliers detection<br>• Outcomes the objects in a particular ordering | • Expects some kind of density decline to find cluster borders<br>• Less sensitive to erroneous data set |
| NC | • Robust to outliers detection<br>• Being parameter-free<br>• Create unique set of neighbours<br>• Geometrically intuitive | • High computational cost<br>• Create fixed of number and neighbours area<br>• Suitable just for small data set |
| *β*-skeleton | • Create changeable of neighbours area<br>• Geometrically intuitive | • Uncertainty of tuning the input parameters<br>• High computational cost |
| RNGN | • Being parameter-free<br>• Robust to outliers detection | • High computational cost<br>• Create fixed number and neighbours area |
| ANGN | • Create flexibly of neighbours area with parameter-angle<br>• Geometrically intuitive<br>• Simple approach | • Uncertainty of tuning the input parameters<br>• high computational cost<br>• Suitable just for small data set |
| AMOEBA | • Low computational cost<br>• Being parameter-free<br>• Less sensitive to noise<br>• Detecting meaningful clusters in data of varying density | • Designed for 2-dimensional data sets |
| SVM | • Robust to outliers detection<br>• Has a regularisation parameter | • Computationally expensive<br>• Runs slowly<br>• Approach lies in choice of the kernel |
| SOM | • Low computational cost<br>• Little memory requirement<br>• Simplicity of the computation<br>• Deterministic reproducible results | • Not so intuitive : neurons close on the map (topological proximity) may be far away in attribute space<br>• Does not behave so gently when using categorical data set, even worse for mixed data set<br>• Relies on a predefined distance in feature space |

**Table 5. The advantages and disadvantages of the discussed categorizing approaches (cont.)**

| | | |
|---|---|---|
| $OPF_{knn}$ | • Processing is fast<br>• Suitable for large data set | • Need to tuning the input parameters<br>• More sensitive to misclassification |
| GMPLC | • Deal with high-dimensional data set<br>• Discover clusters of arbitrary shapes<br>• Processing on nonlinear data sets | • Need to tuning the input parameters<br>• Sensitivity to irrelevant features<br>• Runs slowly |
| CHKNN | • Middle complexity<br>• Processing on linear and nonlinear data sets<br>• Processing on more complex and noisy nonlinear data sets | • Need to tuning the input parameters<br>• Sensitivity to irrelevant features |
| MKAOG | • Less sensitive to noise | • Suitable just for small data set<br>• Suitable just for low-dimensional |
| CDA | • Visualization reduced the total time find connection among data set<br>• Discover insights about the data set by visualization<br>• Ask insightful questions about the data set by visualization<br>• Identify embedded problems in the data set, such as missing, erroneous, or incomplete values<br>• Generate knowledge about the data set by visualization | • Comparing and discovering visualization connection for each data |
| DPC-KNN | • Lack of repetition phase<br>• The centers are extracted from the basic features of the data points | • High computational cost<br>• Suitable just for small data set<br>• Need to define parameter<br>• The number of appropriate clusters obtained from the decision graph may not be equal and comparable with the number of ideal clusters.<br>• Not sensitive to the local geometric features |
| DPC-PCA | • Reducing dimensions of the data set<br>• The centers are extracted from the basic features of the data points | • High computational cost<br>• Suitable just for small data set<br>• Lack of any strategy in dealing with the outlier samples<br>• Would not yield an acceptable performance in overlapping data sets |
| NCAR | • Robust to outliers detection<br>• Geometrically intuitive<br>• Low computational cost | • Suitable just for small data set |

In Table 6 the detailed and comprehensive comparisons of the discussed categorizing approaches are displayed. As Table 6 indicates, the neighbourhood construction algorithms which use neural networks are more suitable for large scale data. Base geometry algorithms do not perform well with large scale data, and problems come about with an increase in the size and dimensions of data sets. Base distance and graph base algorithms usually have medium complexity except for MKAOG which is highly complex. The shortcoming of NC, RNGN, ANGN and $\beta$-skeleton are high time complexity. As such, they are more suitable for examining the data sets in small databases. The shortcoming of NCAR algorithm is suitable for examining the data sets in low dimension and small data sets.

**Table 6. The detailed and comprehensive comparisons of the discussed categorizing approaches**

| Neighbourhood Approaches | Applications | For Large scale data | For high dimensional data | Complexity (time) | References |
|---|---|---|---|---|---|
| *k*-NN | Clustering/classification | No | No | Low | İnkaya et al. 2015, Von Luxburg 2007, Duda et al. 2012, Lu and Fu 1978, Jarvis and Patrick 1973, Tao et al. 2006. |
| DBSCAN | Clustering/classification | Yes | No | Low | Pedrycz 2010, Ankerst et al. 1999, |
| OPTICS | Clustering/classification | Yes | No | Low | Ankerst et al. 1999, Januzaj et al. 2004, Kriegel and Pfeifle 2005 |
| NC | Clustering/classification | No | No | High | İnkaya et al. 2015 |
| *β*-skeleton | Clustering/classification | No | No | High | Langerman 2009, Toussaint 1988, Yang et al. 2016 |
| RNGN | classification | No | No | High | Yang et al. 2016 |
| ANGN | classification | No | No | High | Yang et al. 2016 |
| AMOEBA | Clustering | Yes | No | Low | Estivill-Castro and Lee 2000 |
| SVM | classification | Yes | Yes | High | Güney and Atasoy 2012, Pan et al. 2015 |
| SOM | Clustering | Yes | Yes | Middle | Oja et al. 2003,Roigé et al. 2016 |
| $OPF_{knn}$ | classification | Yes | No | Middle | Papa et al. 2017 |
| GMPLC | Clustering | Yes | Yes | Middle | Wang et al. 2012 |
| CHKNN | Clustering | Yes | Yes | Middle | Qin et al. 2018 |
| MKAOG | classification | No | No | High | Mohammadi et al. 2015 |
| CDA | Visualization/ Clustering | No | No | Middle | Park and Basole 2016 |
| NCAR | Clustering | No | No | Low | Pourbahrami et al. 2018 |

## 7. Conclusions and Future Challenges

In this paper, the issues of finding neighbourhood points and categorizing them were discussed with a focus on the specific challenges. As already discussed the four challenges in categorizing data flow involve changes in neighbourhood and categorizing algorithm conditions necessary for the flow. The approach adopted in the second section focuses on the notion of distance and density among the mentioned data points .However, determining neighbourhood through data points is not accurate enough. Among the proposed algorithms, gradual improvement as a result of gradual conditions was observed. In the third section which was about the algorithms based on geometric structure of the data and graph, different graphs were used for determining neighbourhood points. In the fourth section, the use of artificial neural-network and learning models is discussed. Finally, in the fifth section, the graph-based and density-based algorithms were examined in a structure based on nodes, edges, and the relationship between the nodes. As a general conclusion about neighbourhood construction algorithms discussed in the paper, it can be stated that AMOEBA and NCAR have the lowest time complexity among the clustering algorithms, and $OPF_{knn}$ is less complex among the classification algorithms. Among classification algorithms, MKAOG algorithm exhibited the greatest accuracy in neighbourhood construction over the IRIS data set, and NC showed the highest accuracy over the IRIS data set. AMOEB algorithm is suitable for clustering large data in 2-dimensional data sets. It should be mentioned that most of the algorithms have high complexity as they examine all the neighbourhood relations among the data points, and the presentation of proper algorithms for reducing the complexity is one of the topics to be investigated. To reduce high complexity and plethora of computations in neighbourhood construction approaches, geometric and mathematical algorithms that are not based on parameter specification and do not have fixed neighbourhood areas may be examined in the future.